# Observations of zero-order bandgaps in negative-index photonic crystal superlattices at the near-infrared

S. Kocaman[1], R. Chatterjee[1], N. C. Panoiu[2], J. F. McMillan[1], M. B.Yu[3], R. M. Osgood[1], D. L. Kwong[3], and C. W. Wong[1]

[1]*Columbia University, New York, NY 10027, USA*
[2]*University College London, Torrington Place, London WC1E 7JE, UK*
[3]*The Institute of Microelectronics, Singapore 117685, Singapore*

We present the first observations of zero-$\bar{n}$ bandgaps in photonic crystal superlattices consisting of alternating stacks of negative index photonic crystals and positive index dielectric materials in the near-infrared. Guided by ab initio three-dimensional numerical simulations, the fabricated nanostructured superlattices demonstrate the presence of zero-order gaps in remarkable agreement with theoretical predictions across a range of different superlattice periods and unit cell variations. These volume-averaged zero-index superlattice structures present a new type of photonic band gap, with potential for complete wavefront control for arbitrary phase delay lines and open cavity resonances.

Corresponding authors: cww2104@columbia.edu, sk2927@columbia.edu

Negative index metamaterials (NIM) are material composites having both negative permittivity and permeability, and, consequently, a negative index of refraction. While the existence and basic physical properties of such electromagnetic media have been suggested theoretically almost four decades ago [1], their remarkable properties have only been observed recently, such as with a periodic array of metallic resonators [2-4] or dielectric photonic crystals to emulate an effective negative index [5,6]. Physical properties of such metamaterials have sparked intense interest [7,8], both in their unusual character and in an increasing array of compelling technological applications. In particular, theoretical studies have recently postulated that a superlattice Bragg media of both negative and positive refractive index possesses unique optical properties that cannot be replicated with a periodic positive index material alone [9-11].

One of the most remarkable properties of superlattice structures with alternating negative and positive index materials is *the existence of a new type of photonic band gap*, which cannot exist in media with only positive or only negative indices [9,10]. The formation of this new gap, termed a zero-$\bar{n}$ gap, requires that, at a desired optical frequency, the volume-averaged index is zero and is suggested to possess unusual character such as invariance to incident angles (omnidirectional), invariance to structural disorder and thickness of superlattice layers, with possible observations of unique surface Tamm states and novel resonances [10,11]. Since this gap frequency satisfies the Bragg condition [$k\Lambda=(n\omega/c)\Lambda=m\pi$] for *m*=0, we can also term this a zero-order gap. (*k* and ω are the wave vectors and frequencies respectively, *n* the averaged refractive index, and Λ the superlattice period.)

Here we present the first observations of zero-order gaps in photonic crystal superlattices at near-infrared frequencies. The photonic crystal (PhC) superlattice is shown in Fig. 1, comprising of an alternating negative-positive index superlattice. The negative index layer in the superlattice is based on a negative index PhC we recently demonstrated for subdiffraction imaging based on bound surface states [5]. The designed and fabricated negative index layer possesses a negative index [12] specifically in the frequency range of 0.271 to 0.284 (normalized frequency unit of $\omega a/2\pi c$, where $a$ is the PhC lattice period), or equivalently approximately 1485.1 to 1556.4-nm wavelengths in our measurements [5].

We note that the hexagonal PhC lattice and the superlattice have different symmetry properties and therefore different first Brillouin zones as shown in Fig. 1d [10,13]. The period of the superlattice is defined as $\Lambda = d_1 + d_2$ where $d_1$ and $d_2$ are the thicknesses of the PhC and of the positive index layers in the primary unit cell respectively. We emphasize that the zero-$\bar{n}$ band gaps investigated here, i.e. zero-order band gaps, are *not* due to band-folding back into the first Brillouin zone, as is the case with regular Bragg gaps (intrinsic consequence of periodicity), but rather a consequence of the fact that, at frequencies in which the spatially averaged index of refraction is zero, all supported modes are evanescent (all wave vectors are purely imaginary). Waves reflected from consecutive interfaces situated one $\Lambda$-period apart arrive in-phase at the input facet of the superlattice, leading thus to increased reflectance of the superlattice. Consequently, the transmission through the superlattice decreases and a transmission gap is formed [9,10]. Moreover, we note that our implementation is in the infrared with all-dielectric materials, in contrast to recent metallodielectric centimeter-long structures that operate only at microwave frequencies [14].

The physical origin of the zero-$\bar{n}$ bandgap can be illuminated through the Bloch theorem constraint on the transfer matrix, $T$, of the 1D binary periodic superlattice.

Namely we have $Tr[T(\omega)] = 2\cos\kappa\Lambda$, where $\kappa$ is the wave vector and $Tr$ represents the trace operator. For a double layer unit cell we have:

$$Tr[T(\omega)] = 2\cos\left(\frac{\bar{n}\omega\Lambda}{c}\right) - \left(\frac{Z_1}{Z_2} + \frac{Z_2}{Z_1} - 2\right)\sin\left(\frac{n_1\omega d_1}{c}\right)\sin\left(\frac{n_2\omega d_2}{c}\right) \quad (1)$$

where $n_{1(2)}$, $Z_{1(2)}$ and $d_{1(2)}$ are the refractive index, impedance, and thickness of the first (second) layer and $\bar{n}$ is the average refractive index, $\bar{n}(x) = \frac{1}{\Lambda}\int_0^\Lambda n(x)dx$, respectively. In the general case, when $Z_2 \neq Z_1$, if $\kappa_0\Lambda = \frac{\bar{n}\omega\Lambda}{c} = m\pi$, with $m$ an integer, the relation

$$|Tr[T(\omega)]| = \left|2 + \left(\frac{Z_1}{Z_2} + \frac{Z_2}{Z_1} - 2\right)\sin^2\left(\frac{n_1\omega d_1}{c}\right)\right| \geq 2$$ holds. This relation implies that the

dispersion relation has no real solution for $\kappa$ unless $\frac{n_1\omega d_1}{c}$ is an integer multiple of $\pi$, which is the Bragg condition and thus photonic bandgaps are formed at the corresponding frequencies. However, if the lattice satisfies the unique condition of a spatially averaged zero refractive index ($\bar{n} = 0$), the $Tr[T(\omega)]$ as defined in equation (1) likewise has a magnitude greater than 2, thereby leading to imaginary solutions for all $\kappa$ and thus a spectral gap which does not scale with the lattice constants [9,10].

To investigate the zero-$\bar{n}$ photonic crystal superlattice exactly, we performed full three-dimensional (3D) finite-difference time-domain (FDTD) numerical simulations (with Rsoft FullWAVE). We chose a PhC superlattice with radius-to-lattice ($r/a$) ratio of 0.290, height-to-lattice ($t/a$) ratio of 0.762 and lattice period $a$ of 420 nm, under TM-like polarization (magnetic field parallel to top surface of PhC), in order to create the negative index layer in the superlattice [5]. A tapered input waveguide (240 nm width tapered up to 10.5 µm) is designed such that symmetry constraints preclude

TE-like mode excitation. The PhC longitudinal direction coincides with the Γ-M symmetry axis (z-axis) in order to work with its specific anomalous dispersion band.

We start with three complete superlattices numerically, each with a different superlattice period $\Lambda$: 7, 11 and 15 unit cells in the PhC z-axis such that $d_1 = 3.5\sqrt{3}\,a$, $d_1 = 5.5\sqrt{3}\,a$, and $d_1 = 7.5\sqrt{3}\,a$, respectively. The positive index thickness is then matched such that $\bar{n} = (n_1 d_1 + n_2 d_2)/\Lambda = 0$ [10] while keeping $d_2/d_1$ unchanged, where $n_1$ and $n_2$ are the effective mode indices. We note that, since the zero-$\bar{n}$ gap is formed when the spatially averaged refractive index is zero, it is insensitive to the superlattice period $\Lambda$ as long as the condition of zero-averaged index is satisfied, as postulated analytically. The 3D FDTD computations are repeated for two sets of superlattices with $d_2/d_1$ ratios of 0.746 (design 1) and 0.794 (design 2). For each $d_2/d_1$ ratio, the transmission results show several shifting features with increasing unit cells, except for an *invariant* gap centered at $\omega = 0.276$ in Fig. 2a (design 1) and $\omega = 0.272$ in Fig. 2b (design 2). The shifting features are indicative of conventional Bragg gaps, while the invariant gaps are indicative of the zero-$\bar{n}$ gap. Moreover, with changing $d_2/d_1$ ratio, we confirmed the shift in the zero-$\bar{n}$ gap, since with a different $d_2/d_1$ ratio, there exists another frequency such that the $\bar{n} = (n_1 d_1 + n_2 d_2)/\Lambda = 0$ condition is satisfied (with varying PhC negative index with wavelength). To support our numerical observations, we note that our negative indices are tuned between –1.604 and –1.988 for $\omega$ at 0.295 and 0.276 respectively, as obtained from the band structure (Fig. 1d) [5]. The complete negative index region is also shaded in Fig. 2 for convenience. Any Fabry-Perot reflections from the finite superlattice also do not show up as large (10-dB or more) intensity contrasts. A 5 stack (5 superlattice periods) structure is used for the 7 unit cell

PhC case to obtain a distinct zero-$\bar{n}$ gap, while a 3 stack superlattice is sufficient for the 11 and 15 unit cell cases. The grid size resolution in all our computations is 35 nm.

To further support the nature of these photonic gaps, we present the superlattice Bragg order, $m$ (= $k_o\Lambda/\pi$), in Table 1 for the various cases computed. Particularly, for design 1 (design 2) with $\omega$ at 0.276 (0.272), the average index is determined to be -0.007 (0.001) with corresponding $k_o\Lambda/\pi$ of -0.044 (0.007) – or equivalently, these are the zero-order gaps within numerical discretization certainty. At $\omega$ of 0.283 for designs 1 and 2, the average indices are 0.091 and 0.153 respectively with corresponding $k_o\Lambda/\pi$ of 0.874 and 0.962 – these are the first-order gaps in the superlattice. We emphasize that these zero-order and first-order gaps are *not* from a conventional single positive-index band gap (solely single period photonic crystal) because the regular band gap computed for 15, 30, and 45 units cells (Fig. 2d) is determined to be red-shifted by almost 40 nm in wavelength ($\omega$ of 0.270) from the zero-order gap shown in Fig. 2a. Moreover, for increasing number of superlattice periods (Fig. 2c), the center frequency of the zero-$\bar{n}$ gap is indeed invariant, with an increasing intensity contrast.

Encouraged by these numerical observations, we fabricated actual samples in a silicon-on-insulator substrate with 3, 5 and 8 superperiods (example SEM in Fig. 1a). The negative index PhC layer has thickness $d_l$ of $3.5\sqrt{3}\ a$. The samples are fabricated with a 248-nm lithography, and any remaining fabrication disorder in the PhC was statistically parameterized [15]. The resulting hole radii were 122.207 ± 1.207 nm, with lattice periods of 421.78 ± 1.26 nm (~ 0.003$a$), and hole ellipticities of 1.21 nm ± 0.56 nm. The disorder variations are significantly below ~0.05$a$, preventing any deterioration of the superlattice properties [16].

Transmission measurements were performed on the negative-positive index PhC superlattice, with TM incident polarization from widely-tunable lasers (1480 nm to 1690 nm) coupled to the chip via lensed fibers. Each transmission is averaged over three scans, collected via a 40× objective lens with a slit positioned spatially at the superlattice output facet, and normalized to a regular channel waveguide (240-nm width) transmission. Figure 3a shows the transmission spectrum for design 1 (with $d_2/d_1$ = 0.746). Two distinct gaps centred at 1520 nm ($\omega$ = 0.276 [$\omega a/2\pi c$]) and 1585 nm ($\omega$ = 0.265 [$\omega a/2\pi c$]) were observed. The gap at 1520 nm matches remarkably with our 3D numerical simulations without any fitting. Moreover, the spectral width of the gap matches our simulations (~19 nm). We repeated these transmission measurements for design 2 ($d_2/d_1$ = 0.794), with the transmission results shown in Fig. 3b. A distinct gap is observed at 1543 nm ($\omega$ = 0.272 [$\omega a/2\pi c$]). The measurements show a remarkable correspondence with our numerical predictions without any fitting, further verifying for the first time the existence of a zero-$\bar{n}$ gap in these PhC superlattices.

Figure 3c further confirms the zero-$\bar{n}$ gap with another series of PhC superlattices, each with 7 unit cells ($d_2/d_1$ of 0.746; design 1) but with increasing stack number from 3 to 5 and 8. As observed, the zero-$\bar{n}$ gap intensity contrast increases with increasing number of superperiods. The zero-$\bar{n}$ gap center frequency is invariant with increasing number of superperiods, matching with our predictions. (Slight variations in gap center frequency arise from small variations in the hole radii between the fabricated superlattices.) The regular gap intensity contrast likewise increases with increasing total number of unit cells; however this frequency cannot be the zero-$\bar{n}$ gap because it falls outside the negative index region from our previous measurements [5] and is noted in Fig. 3a and 3b to be the conventional positive-index Bragg gap.

Moreover, in contrast to regular PhC gaps, the zero-$\bar{n}$ gap is surprisingly robust against large nanofabrication-induced symmetric disorder since we require only the volume-averaged zero index [9,10]. Furthermore, we note this first demonstration of the zero-$\bar{n}$ gaps in negative-positive index superlattices can have potential applications such as open cavity resonances [17], delay lines with zero phase differences, curved to planar wavefront transformation, and directional emission of an embedded emitter within the metamaterial [9].

In summary, we demonstrate for the first time the existence of a zero-$\bar{n}$ gap at near-infrared frequencies by using a dielectric PhC superlattice consisting of alternating layers of negative index PhC with positive index homogeneous layers. Our physical measurements on the fabricated nanostructured superlattices with different structural parameters show a remarkable match with our theoretical predictions based on rigorous fully 3D numerical simulations. The presence of the zero-order gap is verified for structures of different superlattice ratios ($d_2/d_1$), different number of unit cells, and different number of superperiods, along with ab initio predictions of the gap order, and can have implications for arbitrary phase delay lines for communications and open cavity resonances.

S. K., R. C, J. F. M. and C. W. W. were supported by a 2008 NSF CAREER Award (ECCS-0747787), a 2007 DARPA Young Faculty Award (W911NF-07-1-0175), and the New York State Foundation for Science, Technology and Innovation. N. C. P. and R. M. O were supported by NSF (ECCS-0523386).

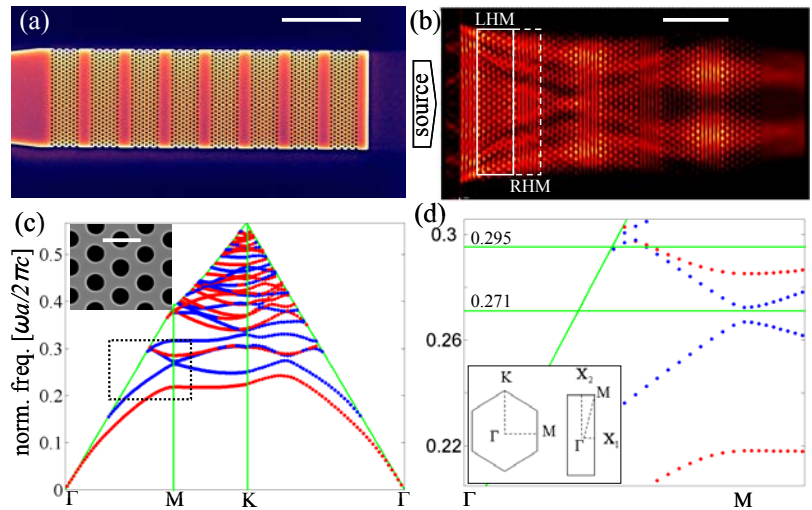

**Figure 1.** (Color online) (a) SEM of a fabricated sample with 8 stacks, whose PhC slab layer has a length of $d_1 = 3.5\sqrt{3}\,a$. Scale bar: 5 μm. (b) A time-averaged steady-state numerically calculated of the field intensity distribution, $|E|^2$, corresponding to a propagating mode ($\lambda = 1550$ nm). Scale bar: 5 μm. (c) Calculated photonic band structure of the fabricated PhC slab waveguide with $r = 0.290a$ and $t = 0.762a$ ($a = 420$ nm). The TM-like (TE-like) photonic bands are depicted in blue (red). Inset: SEM of the PhC region of the fabricated superlattice. Scale bar: 500 nm. (d) A zoom-in of the spectral domain corresponding to our experiments. Inset: The first Brillouin zones of the hexagonal PhC and the PhC superlattice.

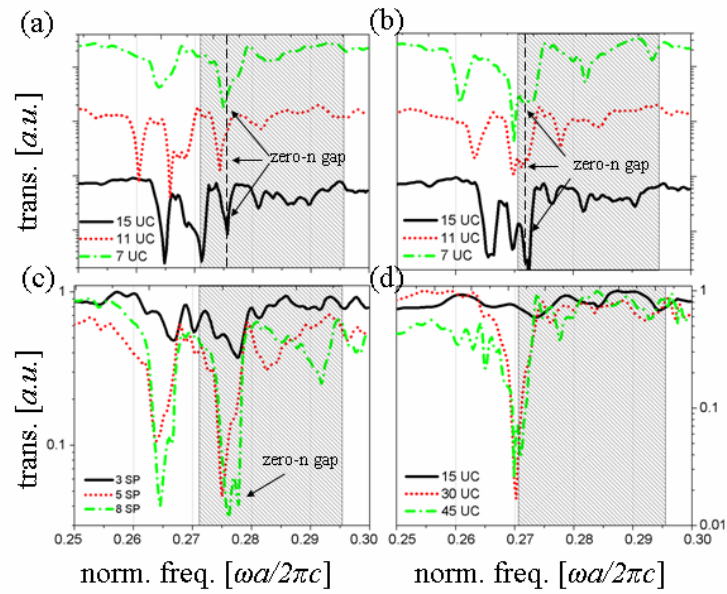

**Figure 2.** (Color online) (a) Transmission highlighting the invariant zero-$\bar{n}$ gaps for a superlattice with $d_2/d_1 = 0.746$ (design 1), containing 7, 11 and 15 unit cells in each PhC slab. 3 superperiods in the PhC superlattice are used for the case of PhC layers with 11 and 15 unit cells, and 5 superperiods for the case of PhC layers with 7 unit cells. The shaded region illustrates the negative index range, in order to observe the zero-$\bar{n}$ gap. (b) The same as in (a), but $d_2/d_1 = 0.794$ (design 2). The shifting gaps are the conventional Bragg gaps. The transmission plots are offset vertically for viewing clarity. (c) Transmission for a superlattice with $d_2/d_1 = 0.794$, containing 3, 5 and 8 superperiods. Each PhC layer contains 7 unit cells. (d) Transmission through a regular PhC slab (non-superlattice) containing 15, 30 and 45 unit cells, illustrating a regular Bragg gap. All designs are obtained through full 3D FDTD numerical simulations. In all plots, the shaded region illustrates the negative index region.

**Table 1.** Average index of corresponding gaps and the gap order in superlattice.

| Figure | Gap Freq. | Eff. Index PhC | Eff. Index Slab | Average Index, $\bar{n}$ | $k_o\Lambda/\pi$ | Gap Order |
|---|---|---|---|---|---|---|
| Fig. 2a (7 UC PhC) | 0.276 | -1.988 | 2.648 | -0.007 | -0.044 | 0 |
| Fig. 2a (11 UC PhC) | 0.283 | -1.844 | 2.684 | 0.091 | 0.874 | 1 |
| Fig. 2b (7 UC PhC) | 0.272 | -2.080 | 2.622 | 0.001 | 0.007 | 0 |
| Fig. 2b (7 UC PhC) | 0.283 | -1.856 | 2.683 | 0.153 | 0.962 | 1 |

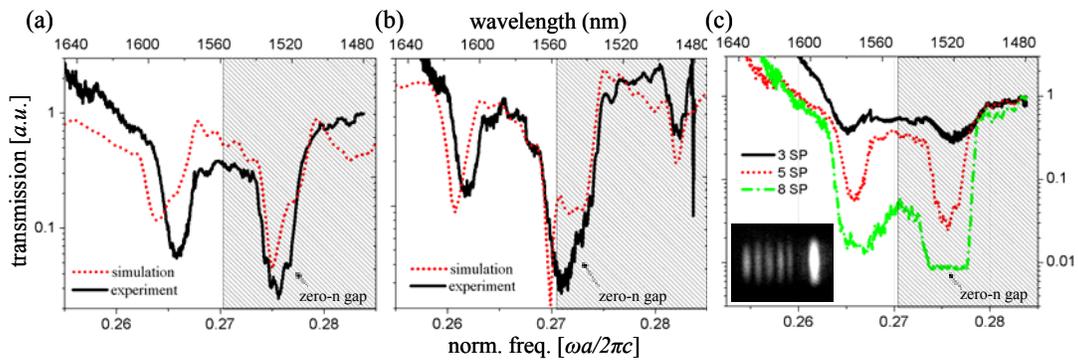

**Figure 3.** (Color online) (a) Measured transmission for a superlattice with $d_2/d_1$= 0.746, with 7 unit cells in the PhC layers and 5 superperiods; for comparison, results of numerical simulations are also shown. (b) The same as in a), but for a superlattice with 0.794. (c) Measured transmission for a superlattice with $d_2/d_1$= 0.746, with 3, 5 and 8 superperiods and 7 unit cells in the PhC layers. Both gaps become deeper as the number of stacks increases. Inset: example of near-infrared top-view image of 3 superperiods, under transmission measurement at 1550 nm. In all plots, the shaded region illustrates the negative index region.